\documentclass[aps,twocolumn,showpacs,amsmath,amssymb]{revtex4}
\usepackage[dvips]{graphics}
\def\imo{i}
\def\a{\widetilde{\alpha}}

\begin{document}
\title{Eikonal instability of Gauss-Bonnet-(anti-)de Sitter black holes}
\author{R. A. Konoplya}
%\email{roman.konoplya@gmail.com}
\affiliation{Theoretical Astrophysics, Eberhard-Karls University of T\"ubingen, T\"ubingen 72076, Germany}
\author{A. Zhidenko}
%\email{olexandr.zhydenko@ufabc.edu.br}
\affiliation{Centro de Matem\'atica, Computa\c{c}\~ao e Cogni\c{c}\~ao, Universidade Federal do ABC (UFABC),\\
  Rua Aboli\c{c}\~ao, CEP: 09210-180, Santo Andr\'e, SP, Brazil}
%\date{\today}

\begin{abstract}
Here we have shown that asymptotically anti-de Sitter (AdS) black holes in the Einstein-Gauss-Bonnet (GB) theory are unstable under linear perturbations of space-time in some region of parameters. This (\emph{eikonal}) instability develops at high multipole numbers. We found the exact parametric regions of the eikonal instability and extended this consideration to asymptotically flat and de Sitter cases. The approach to the threshold of instability is driven by purely imaginary quasinormal modes, which are similar to those found recently in \cite{Grozdanov:2016vgg} for the higher curvature corrected black hole with the planar horizon. The found instability may indicate limits of holographic applicability of the GB-AdS backgrounds. Recently, through the analysis of critical behavior in AdS space-time in the presence of Gauss-Bonnet term, it was shown \cite{Deppe:2014oua} that that, if the total energy content of the AdS space-time is small, then no black holes can be formed with mass less than some critical value. A similar mass gap was also found when considering collapse of mass shells in asymptotically flat Gauss-Bonnet theories \cite{Frolov:2015bta}. The found instability of \emph{all sufficiently small} Einstein-Gauss-Bonnet-AdS, -dS and asymptotically flat black holes may explain the existing mass gaps in their formation.
\end{abstract}

%\pacs{04.30.Nk,04.50.+h}
\maketitle

\section{Introduction}

Perturbations of asymptotically AdS black holes can be interpreted in terms of the renown guage/gravity duality and used for description of quark-gluon plasma \cite{Son:2007vk}. At late times after the collision of particles, hydrodynamics provides a good effective description of the system. Though it is not appropriate for analysis of earlier evolution of plasma far from equilibrium. Equilibration and thermalization of the quantum system is related to the formation of the black-hole horizon and its equilibration on the gravity's side of the duality. The least damped proper oscillation frequencies of the black-holes' \emph{quasinormal spectra} \cite{Konoplya:2011qq} determine the timescale for  relaxation of perturbations \cite{Chesler:2008hg}.

One of the main current problems of such a holographic description is connected with the fact that the dual field theory has an infinite 't Hooft coupling $\lambda$ \cite{Son:2007vk}. However, the regime of finite $\lambda$ can be achieved through higher curvature corrections to the gravitational action \cite{Waeber:2015oka,Grozdanov:2016vgg}. Higher curvature corrected theories of gravity are also important alternatives to the Einstein gravity, as they appear in the low-energy limit of the heterotic string theory. The Gauss-Bonnet (second order in curvature) correction to the Einstein action is the dominant one;
\begin{equation}\label{gbg3}
  \mathcal{L}=-2\Lambda+R+\frac{\alpha}{2}(R_{\mu\nu\lambda\sigma}R^{\mu\nu\lambda\sigma}-4\,R_{\mu\nu}R^{\mu\nu}+R^2),
\end{equation}
where $\alpha$ is a coupling constant, which need not be positive in the context of holography, but it is usually implied that $\alpha=1/2\pi l_s^2$ in the higher-dimensional gravity. The onset of instability on the AdS side indicates the phase transition in the dual field theory, what may signify the limits of holographic applicability of the gravitational background.  Therefore, the natural questions are whether the Gauss-Bonnet-AdS black holes have regions of instability and what are their quasinormal spectra in the stability sector.

The study of quasinormal modes of black holes in the Gauss-Bonnet theory started from computations of the dominant quasinormal modes for a test scalar field \cite{Konoplya:2004xx} and further was extended to gravitational perturbations in the asymptotically flat \cite{Konoplya:2008ix} and de Sitter \cite{Cuyubamba:2016cug} worlds. Then, the WKB estimations were performed for asymptotically flat Lovelock black holes \cite{Yoshida:2015vua}. Nevertheless, no analysis of gravitational stability or quasinormal spectra was performed for spherically symmetric Einstein-Gauss-Bonnet-anti-de Sitter black holes.

Asymptotically flat black holes in higher curvature theories have a number of peculiar features. They are characterized by anomalously weaker Hawking evaporation rate \cite{Konoplya:2010vz}. The graviton experiences time advance at a nonzero Gauss-Bonnet correction \cite{Camanho:2014apa}. Unlike Schwarzschild  space-time, small spherically-symmetric and asymptotically flat black holes in the Gauss-Bonnet and Lovelock theories are unstable \cite{Gleiser:2005ra,Dotti:2005sq,Takahashi:2010gz,Gonzalez:2017gwa,Takahashi:2012np,Gannouji:2013eka} and the instability is unusual as, counterintuitively, it develops at higher multipole numbers $\ell$ \cite{Konoplya:2008ix}.

After all, an essential motivation for our study is related to the qualitatively different nonlinear behavior of asymptotically AdS space-times. The empty AdS space-time is linearly stable, but it is unstable against nonlinear perturbations in the Einstein gravity \cite{Bizon}. A way to stabilize the AdS space-time is known: collapse of the perturbation energy, what ends up with a stable AdS black hole in Einstein theory. In \cite{Deppe:2014oua} it was shown that adding the Gauss-Bonnet correction prevents formation of the black hole with small mass, what was interpreted as the stability gap of AdS space-times in GB theory \cite{Deppe:2014oua}. Remarkably, when considering collapse of thin massive shells in the asymptotically flat space-time and higher derivatives in the theory \cite{Frolov:2015bta}, the mass gap in the formation of a black hole was also found, not allowing the formation of black holes of sufficiently small mass. Thus, this phenomenon seems does not depend on the asymptotic behaviour of space-time.

Here we show that Gauss-Bonnet-AdS black holes suffer from a special (eikonal \cite{Cuyubamba:2016cug}) kind of gravitational instability. By finding \emph{exact parametric regions} of this instability and the black-hole quasinormal spectrum we try to understand both of the above aspects of AdS space-time in GB theory: its usefulness for holographic description and relation to the known nonlinear (in)stabilities in AdS space-time.

The paper is organized as follows: In Sec.~\ref{sec:GBdS}~and~\ref{sec:perturbations} we briefly relate the basic information on the black-hole metric
and perturbation equations respectively. Sec.~\ref{sec:eikonal} is devoted to deduction of the exact parametric region of the eikonal instabilities. In Sec.~\ref{sec:QNMs} we discuss the details of quasinormal spectra of the Gauss-Bonnet-AdS case and possible consequences of the observed here instability for holography.
Sec.~\ref{sec:massgap} discusses the found instability in the context of recently found mass gaps in the formation of small black holes in higher curvature corrected theories.

\section{Gauss-Bonnet-(anti-)de Sitter black hole}\label{sec:GBdS}

A black hole solution in the $D$-dimensional Einstein-Gauss-Bonnet theory (\ref{gbg3}) has the form \cite{Boulware:1985wk}
\begin{equation}\label{gbg4}
  ds^2=-f(r)dt^2+\frac{1}{f(r)}dr^2 + r^2\,d\Omega_n^2
\end{equation}
where $d\Omega_n^2$ is a $(n=D-2)$-dimensional sphere, and
\begin{equation}\label{fdef}
f(r)=1-r^2\,\psi(r),
\end{equation}
where $\psi(r)$ satisfies
\begin{equation}\label{Wdef}
W[\psi]\equiv\frac{n}{2}\psi(1 + \a\psi) - \frac{\Lambda}{n + 1} = \frac{\mu}{r^{n + 1}}\,.
\end{equation}
Here
$\a\equiv \alpha\dfrac{(n - 1) (n - 2)}{2},$
and $\mu$ is a constant, proportional to mass; the solution of (\ref{Wdef}) which goes over into the corresponding GR-solutions when $\a\to0$ has the form
\begin{equation}\label{psidef}
  \psi(r)=\frac{4\left(\frac{\mu}{r^{n+1}}+\frac{\Lambda}{n+1}\right)}{n+\sqrt{n^2+8\a n\left(\frac{\mu}{r^{n+1}}+\frac{\Lambda}{n+1}\right)}}.
\end{equation}

In order to introduce the parametric space for the black holes we first notice that, for a given negative value of $\a$, there is a lower bound on the mass parameter
\begin{equation}\label{massbound}
\mu>\frac{n\,(-2\a)^{(n-1)/2}}{4}\left(1+\frac{8\a\Lambda}{n(n+1)}\right),
\end{equation}
for which there is an event horizon $r_H>0$. It follows from (\ref{massbound}) that for $\a<0$,
\begin{equation}\label{horizonbound}
r_H^2>-2\a=-(n-1)(n-2)\alpha.
\end{equation}
Further we shall show that the right-hand side of (\ref{massbound}) is positive for any $\Lambda$, implying that the black hole exists only for positive values of the asymptotic mass.

In order to measure all quantities in the units of the same dimension we express $\mu$ as a function of $r_H$:
\begin{equation}\label{massdef}
  \mu=\frac{n\,r_H^{n-1}}{2}\left(1+\frac{\a}{r_H^2}-\frac{2\Lambda  r_H^2}{n(n+1)}\right).
\end{equation}

For the AdS case we measure $\Lambda$ in units of the AdS radius $R$, defined by relation
\begin{equation}\label{Rdef}
\psi(r\to\infty)=-\frac{1}{R^{2}}.
\end{equation}
Then,
\begin{equation}\label{AdSlambda}
  \Lambda=-\frac{n(n+1)}{2R^2}\left(1-\frac{\a}{R^2}\right),
\end{equation}
implying that
\begin{equation}\label{alphabound}
\a<R^2.
\end{equation}

It turns out that, when $R^2/2<\a<R^2$, the solution (\ref{psidef}) does not satisfy (\ref{Rdef}) and describes a black hole, which is identical to the one with $\a<R^2/2$ after some rescaling of parameters. Therefore, here we shall consider black holes with $\a \leq R^2/2$.

Substituting (\ref{AdSlambda}) into (\ref{massdef}) we find that
$$\mu=\frac{nr_H^{n-3}(R^2+r_H^2)}{2R^4}\Biggr(R^2r_H^2+\a(R^2-r_H^2)\Biggr),$$
which is positive, when inequalities (\ref{horizonbound}) and (\ref{alphabound}) are satisfied.

In the de Sitter space-times the span of the spatial coordinate is limited by the cosmological horizon $r_C>r_H$, which we use in order to parametrize the cosmological constant as
\begin{equation}\label{cosmological}
  \Lambda=\frac{n(n+1)}{2}\Biggr(\frac{r_C^{n-1}-r_H^{n-1}}{r_C^{n+1}-r_H^{n+1}}+\a\frac{r_C^{n-3}-r_H^{n-3}}{r_C^{n+1}-r_H^{n+1}}\Biggr).
\end{equation}

In the limit $r_C\to r_H$ we obtain the extremal value of the cosmological constant, which is given as follows
\begin{equation}\label{nr2}
  \Lambda_{extr}=\frac{n(n-1)}{2r_H^2}+\frac{n(n-3)\a}{2r_H^4}.
\end{equation}
Limit $r_C\to\infty$ corresponds to the asymptotically flat space-time ($\Lambda=0$).

Note, that, due to (\ref{horizonbound}), we have
$$\a>-\frac{r_H^2}{2}>-r_H^2\frac{n-1}{n-3}>-\frac{r_C^{n-1}-r_H^{n-1}}{r_C^{n-3}-r_H^{n-3}}.$$
Hence the cosmological constant, defined by (\ref{cosmological}), is always nonnegative.

Substituting (\ref{cosmological}) into (\ref{massdef}) we find that the mass parameter is always positive in the asymptotically de Sitter case
$$\mu=\frac{nr_C^{n-3}r_H^{n-3}}{2}\cdot\frac{r_C^2-r_H^2}{r_C^{n+1}-r_H^{n+1}}\Biggr(r_C^2r_H^2+\a(r_C^2+r_H^2)\Biggr)>0$$
for $r_C^2>r_H^2>-2\a$.

\section{Perturbation equations}\label{sec:perturbations}
The perturbation equations can be reduced to the second-order master differential equations \cite{Takahashi:2010ye}
\begin{equation}\label{gp9}
\left(\frac{\partial^2}{\partial t^2}-\frac{\partial^2}{\partial r_*^2}+V_i(r_*)\right)\Psi(t,r_*)=0,
\end{equation}
where $r_*$ is the tortoise coordinate,
\begin{equation}
dr_*\equiv \frac{dr}{f(r)}=\frac{dr}{1-r^2\psi(r)},
\end{equation}
and $i$ stands for $t$ (\emph{tensor}), $v$ (\emph{vector}), and $s$ (\emph{scalar}) perturbations.
The explicit forms of the effective potentials $V_s(r)$, $V_v(r)$, and $V_t(r)$ are given by \cite{Cuyubamba:2016cug}
\begin{eqnarray}%\nonumber
V_t(r)&=&\frac{\ell(\ell+n-1)f(r)T''(r)}{(n-2)rT'(r)}+\frac{1}{R(r)}\frac{d^2R(r)}{dr_*^2},\\\label{potentials}\nonumber
V_v(r)&=&\frac{(\ell-1)(\ell+n)f(r)T'(r)}{(n-1)rT(r)}+R(r)\frac{d^2}{dr_*^2}\Biggr(\frac{1}{R(r)}\Biggr),\\\nonumber
V_s(r)&=&\frac{2\ell(\ell+n-1)f(r)P'(r)}{nrP(r)}+\frac{P(r)}{r}\frac{d^2}{dr_*^2}\left(\frac{r}{P(r)}\right),
\end{eqnarray}
where $\ell=2,3,4,\ldots$ is the multipole number and
\begin{eqnarray}
T(r)&=& r^{n-1}\frac{dW}{d\psi}=\frac{nr^{n-1}}{2}\Biggr(1+2\a\psi(r)\Biggr),\\\nonumber
R(r)&=&r\sqrt{T'(r)},\\\nonumber
P(r)&=&\frac{2(\ell-1)(\ell+n)-nr^3\psi'(r)}{\sqrt{T'(r)}}T(r).
\end{eqnarray}

From (\ref{Wdef}) we find that $W$ depends on $T^2$, which is positive by definition,
$$W(r)=\frac{n}{2\a}\left(\left(\frac{T(r)}{nr^{n-1}}\right)^2-\frac{1}{4}-\frac{2\a\Lambda}{n(n+1)}\right)=\frac{\mu}{r^{n + 1}}.$$
Therefore, we have
\begin{equation}\label{Tdef}
T^2=n^2r^{2n-2}\left(\frac{1}{4}+\frac{2\a\mu}{nr^{n + 1}}+\frac{2\a\Lambda}{n(n+1)}\right)\,,
\end{equation}
implying that
\begin{equation}\label{neq}
\frac{1}{4}+\frac{2\a\mu}{nr^{n + 1}}+\frac{2\a\Lambda}{n(n+1)}\geq0.
\end{equation}

We also further use the following inequality
\begin{equation}\label{pars}
1+\frac{8\a\Lambda}{n(n+1)}\geq0,
\end{equation}
which follows from (\ref{neq}) for $\a<0$. When $\a>0$ and $\Lambda<0$ we substitute (\ref{AdSlambda}) and observe that the right-hand side of (\ref{pars}) is the perfect square.

\section{Eikonal instability}\label{sec:eikonal}
Intuitively, larger multipole numbers $\ell$ should lead to a higher barrier of the effective potential and thus stabilize the system. This is not the case for theories with higher curvature corrections for which larger $\ell$ correspond to deeper negative gaps near the event horizon. This phenomenon was found for asymptotically flat and de Sitter GB black holes and called \emph{eikonal instability} \cite{Cuyubamba:2016cug}. It occurs when the dominant term in (\ref{potentials}) has a negative gap, so that the potential is negatively dominant for sufficiently large $\ell$ \cite{Takahashi:2010gz}.

For large $\ell$ from \cite{Takahashi:2010gz}, we have
\begin{eqnarray}\nonumber
V_t(r)&=&\frac{\ell^2f(r)T''(r)}{(n-2)rT'(r)}+{\cal O}(\ell),\\\label{dominant}
V_v(r)&=&\frac{\ell^2f(r)T'(r)}{(n-1)rT(r)}+{\cal O}(\ell),\\\nonumber
V_s(r)&=&\frac{\ell^2f(r)(2T'(r)^2-T(r)T''(r))}{nrT'(r)T(r)}+{\cal O}(\ell).
\end{eqnarray}
Analytical proof of sufficiency of a negative gap in (\ref{dominant}) for AdS  black-hole instability is similar to that for the asymptotically flat case \cite{Takahashi:2010gz}.

\emph{Vector channel.}
From (\ref{Tdef}) and (\ref{neq}) it follows that
$$\frac{f(r)T'(r)}{(n-1)rT(r)}\geq\frac{f(r)r^{2n-4}n}{8(n-1)T(r)^2}\left(n(n+1)+8\a\Lambda\right),$$
which is nonnegative due to (\ref{pars}). Hence there is no eikonal instability in the vector mode.

\begin{figure*}
\resizebox{\linewidth}{!}{\includegraphics*{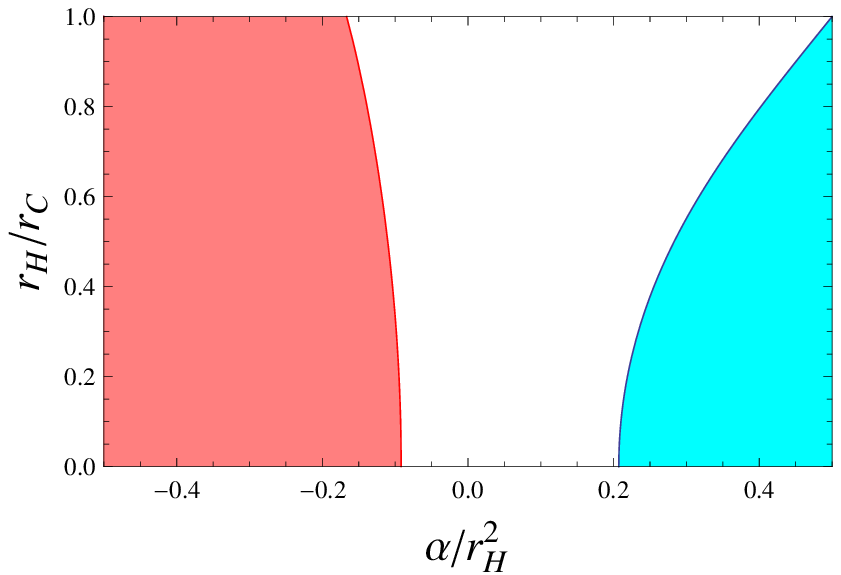}\includegraphics*{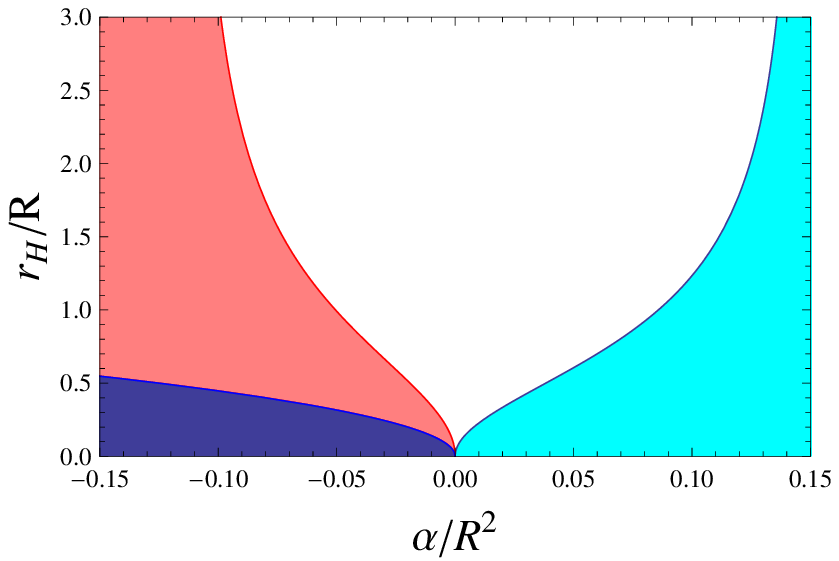}}
\caption{Parametric regions of the eikonal instability for tensor-type perturbations (red, middle) and scalar-type perturbations (cyan) of $D=5$ ($n=3$) Gauss-Bonnet-de Sitter black holes (left panel) and Gauss-Bonnet-anti-de Sitter black holes (right panel). The dark lower left corner of the right panel corresponds to the region given by (\ref{horizonbound}), in which no event horizon exists.}\label{fig:n3eikonal}
\end{figure*}

\emph{Scalar channel.} From the above it follows that the eikonal instability of scalar-type perturbations exists if and only if \cite{Takahashi:2010gz}
$$M(r)\equiv r^2T(r)^2(2T'(r)^2-T(r)T''(r))<0.$$
We are in position to prove that the eikonal instability in the scalar channel exists only for $n=3$ ($D=5$). First, using (\ref{Tdef}), we find that
\begin{eqnarray}
\nonumber M(r)=\frac{n^2r^{4n-4}}{16}\Biggr(n^2(n-7)\left(1+\frac{8\a\Lambda}{n(n+1)}\right)\left(\frac{4\a\mu}{r^{n+1}}\right)+\\\nonumber
n^3(n-1)\left(1+\frac{8\a\Lambda}{n(n+1)}\right)^2+(n-1)(n-3)\left(\frac{4\a\mu}{r^{n+1}}\right)^2\Biggr).
\end{eqnarray}
Then, using (\ref{neq}), for $n\geq7$ ($D\geq9$) we find that
$$M(r)\geq\frac{3n^2r^{4n-4}}{64}\left(n(n+1)+8\a\Lambda\right)^2\geq0.$$

For $n=4,5,6$ one can check that $M(r)\geq0$ for an arbitrary value of $r$. For instance, for $n=4$ ($D=6$)
$$M(r)=48r^2\left(2r^5+\frac{4}{5}r^5\a\Lambda-\a\mu\right)^2\geq0.$$

For $n=3$ ($D=5$) the negative gap in $M$ for $r>r_H$ exists if $M(r_H)<0$ implying that
\begin{equation}\label{n3s}
(3+2\a\Lambda)(3r_H^4+4\a\Lambda r_H^2-12\a r_H^2-12\a^2)<0.
\end{equation}

From (\ref{n3s}) one can see that in the special case $3+2\a\Lambda=0$, inequality (\ref{n3s}) is not satisfied. In the de Sitter space when $\Lambda=-\frac{3}{2\a}$, the lower bound for a black-hole size (\ref{horizonbound}) does not allow for black holes, which are thereby too large to be embedded into the de Sitter space. In the asymptotically anti-de Sitter space-time it corresponds to $\a=\frac{R^2}{2}$ ($\Lambda=-\frac{3}{R^2}$) for which (\ref{fdef}) takes the form
$$f(r)=\frac{r^2}{R^2}+1-\sqrt{\frac{4\mu}{3R^2}}=\frac{r^2-r_H^2}{R^2}.$$
In this case $T^2=6\a\mu=3R^2\mu$ is a constant and the above analysis does not apply. However, the corresponding solution also suffers from the eikonal instability for the scalar-type gravitational perturbations \cite{Gonzalez:2017gwa}.

In the asymptotically de Sitter space-time we use (\ref{cosmological}) to rewrite (\ref{n3s}) as
$$(4\a + r_H^2 + r_C^2) \left(r_H^4-4\a r_H^2\frac{r_C^2-r_H^2}{r_C^2+r_H^2}-\a^2\right)<0,$$
implying that the eikonal instability occurs for (see Fig.~\ref{fig:n3eikonal})
$$\a>r_H^2\frac{\sqrt{2(r_C^4+r_H^4)}-(r_C^2-r_H^2)}{2(r_C^2+r_H^2)}\,,$$
which agrees with the numerical data \cite{Cuyubamba:2016cug}.

In the limit of the asymptotically flat space $r_C\to\infty$ we have the eikonal instability for
$$\a>\frac{\sqrt{2}-1}{2}r_H^2\approx0.2071r_H^2,$$
coinciding with the analytical value found in \cite{Beroiz:2007gp} and numerical one of \cite{Konoplya:2008ix}.

In the asymptotically anti-de Sitter space-time we use (\ref{AdSlambda}) to rewrite (\ref{n3s}) as
\begin{eqnarray}
\left(\a-\frac{R^2}{2}\right)^2\Biggr(r_H^4-4\a r_H^2\left(1+\frac{2r_H^2}{R^2}\right)&&\\\nonumber-4\a^2\left(1-\frac{2r_H^2}{R^2}\right)\Biggr)&<&0,
\end{eqnarray}
leading to the eikonal instability for (see Fig.~\ref{fig:n3eikonal})
\begin{equation}\label{uplimit}
\a>\frac{R^2r_H^2}{2}\frac{\sqrt{2}-1}{\sqrt{2}r_H^2+R^2}.
\end{equation}
In particular, for $\a>\frac{\sqrt{2}-1}{2\sqrt{2}}R^2\approx 0.1464466 R^2$ all black holes are unstable.

\begin{figure*}
\resizebox{\linewidth}{!}{\includegraphics*{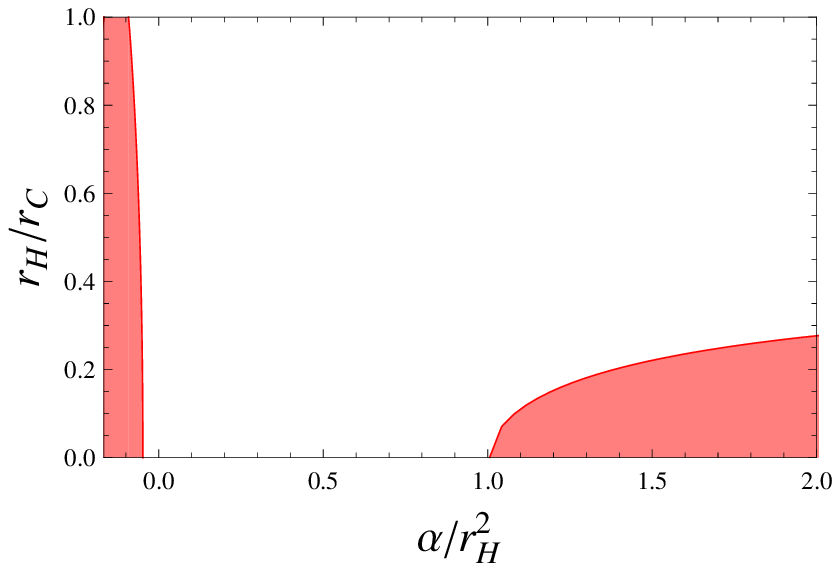}\includegraphics*{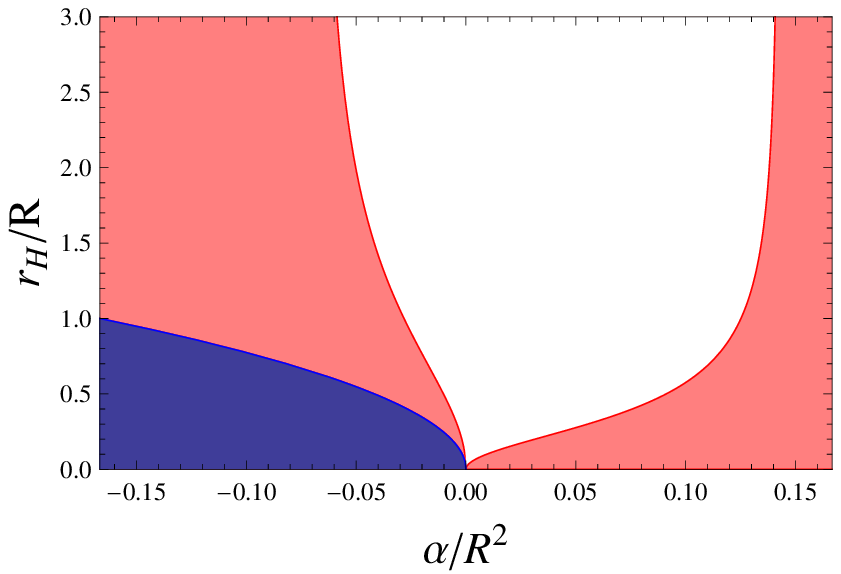}}
\caption{Parametric regions of the eikonal instability for tensor-type perturbations of $D=6$ ($n=4$) Gauss-Bonnet-de Sitter black holes (left panel) and Gauss-Bonnet-anti-de Sitter black holes (right panel). The dark lower left corner of the right panel corresponds to the the region given by (\ref{horizonbound}), in which no event horizon exists.}\label{fig:n4eikonal}
\end{figure*}

\emph{Tensor channel.} Taking into account that $T'(r)T(r)>0$ it was shown in \cite{Takahashi:2010gz} that the eikonal instability of tensor-type perturbations exists if and only if
$$T''(r)T(r)\equiv\frac{N(r)}{r^2T(r)^2}<0.$$
From (\ref{Tdef}) we find the explicit form of $N(r)$,
\begin{eqnarray}
\nonumber N(r)&=&\frac{n^4r^{4n-4}}{16}\Biggr((n-3)(n-5)\left(\frac{4\a\mu}{r^{n+1}}\right)^2\\\nonumber&&
+n^2(n-1)(n-2)\left(1+\frac{8\a\Lambda}{n(n+1)}\right)^2\\\nonumber&&
+3n(n^2-3n+4)\left(1+\frac{8\a\Lambda}{n(n+1)}\right)\left(\frac{4\a\mu}{nr^{n+1}}\right)\Biggr).
\end{eqnarray}
It is obvious that there is no eikonal instability in the tensor-type perturbations for $\a\geq0$, unless $n=4$ \cite{Dotti:2005sq}. Yet, if one considers negative values of $\a$ there are parametric regions of the eikonal instability for any $n$. It is interesting to note that, since by definition
$$M(r)+N(r)=2T'(r)^2T(r)^2r^2\geq0,$$
the parametric regions of eikonal instability for scalar and tensor sectors have no intersections (see Fig.~\ref{fig:n3eikonal}).

For $n=3$ ($D=5$, $\a=\alpha$) the negative gap in $N$ for $r>r_H$ exists if $N(r_H)<0$ implying
\begin{equation}\label{n3t}
(3+2\a\Lambda)(3r_H^4-4\a\Lambda r_H^4+36\a r_H^2+36\a^2)<0.
\end{equation}

In the asymptotically de Sitter space-time, using (\ref{cosmological}), we rewrite (\ref{n3t}) as
$$(4\a + r_H^2 + r_C^2) \left(r_H^4+4\a r_H^2\frac{3r_C^2+r_H^2}{r_C^2+r_H^2}+12\a^2\right)<0,$$
implying that the eikonal instability occurs for (see Fig.~\ref{fig:n3eikonal})
\begin{equation}
\a<-r_H^2\frac{3r_C^2+r_H^2-\sqrt{6r_C^4-2r_H^4}}{6(r_C^2+r_H^2)}\,.
\end{equation}

In the asymptotically anti-de Sitter space-time, using (\ref{AdSlambda}), we rewrite (\ref{n3t}) as
\begin{eqnarray}\nonumber
\left(\a-\frac{R^2}{2}\right)^2\Biggr(r_H^4+4\a r_H^2\left(3+\frac{2r_H^2}{R^2}\right)&&\\\nonumber+4\a^2\left(3-\frac{2r_H^2}{R^2}\right)\Biggr)&<&0,
\end{eqnarray}
leading to the eikonal instability for (see Fig.~\ref{fig:n3eikonal})
\begin{equation}\label{lowlimit}
\a<-\frac{R^2r_H^2}{2}\frac{\sqrt{3}-\sqrt{2}}{\sqrt{3}R^2+\sqrt{2}r_H^2}.
\end{equation}

For $n=4$ ($D=6$) the negative gap in $N$ exists for
\begin{eqnarray}\label{n4t}
&&4 \a^4 + 8 \a^3 r_H^2 - 44 \a^2 r_H^4 - 48 \a r_H^6 \\\nonumber&&\qquad
 -6 r_H^8 - 20 \a^3 r_H^4 \Lambda - 20 \a^2 r_H^6 \Lambda + \a^2 r_H^8 \Lambda^2>0.
\end{eqnarray}

Substitutions of (\ref{cosmological}) and (\ref{AdSlambda}) into (\ref{n4t}) lead to cumbersome inequations of degree four with respect to $\a$. The parametric regions of the eikonal instability are presented on Fig.~\ref{fig:n4eikonal} (note that $\a=3\alpha$). The left panel, namely the case $\alpha>0$ and $\Lambda>0$, agrees with \cite{Cuyubamba:2016cug}, where this instability was studied numerically. In particular, for the flat space-time we obtain the region free of the eikonal instability
$$-\frac{1-\sqrt{25-10\sqrt{6}}}{6}r_H^2\leq\alpha\leq\frac{\sqrt{25+10\sqrt{6}}-1}{6}r_H^2.$$
The positive bound agrees with the values obtained theoretically \cite{Dotti:2005sq} and numerically \cite{Konoplya:2008ix}: $\alpha\approx1.006r_H^2$.

Notice that in addition to the eikonal instability, considered here, Einstein-Gauss-Bonnet-de Sitter black holes have the instability developed at the lowest multipole number, which is related to the nonzero cosmological constant \footnote{Apparently, a similar instability takes place for the higher-dimensional Reissner-Nordstr\"om-de Sitter black holes without Gauss-Bonnet term \cite{Konoplya:2008au}.}. Thus, the resultant region of instability of the Gauss-Bonnet-de Sitter black hole can be found as a union of the eikonal and $\Lambda$-term induced regions of instabilities \cite{Cuyubamba:2016cug}.

\section{Eikonal instability, quasinormal frequencies and holography}\label{sec:QNMs}

When gravitational instability occurs at some values of the parameters of the system, the regime of linear perturbations breaks down, because the unbounded exponential growth of the perturbation is outside the linear approximation. For the GB theory there is another reason for the breakdown of the perturbation formalism: the nonhyperbolicity of the higher curvature theories \cite{Reall:2014pwa} at some values of space-time parameters.

Using the shooting method, essentially in the same way as in \cite{Konoplya:2008rq}, we find eigenfrequencies of the master wave equation with an effective potential for each type of perturbation (scalar, vector, and tensor) at the so-called quasinormal boundary conditions, which require purely incoming waves at the black-hole horizon and the Dirichlet condition at infinity.

\begin{figure}
\resizebox{\linewidth}{!}{\includegraphics*{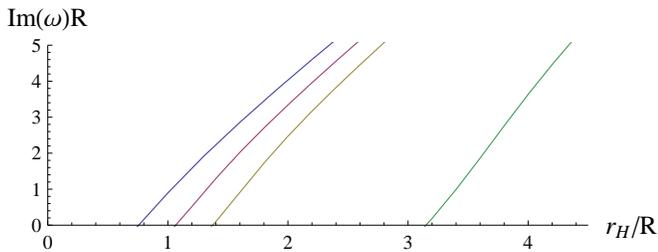}}
\caption{The imaginary part of $\omega$ at $\ell=2, 3, 4, 10$ (from left to right) for $\alpha = 0.4R^2$, $D=5$, scalar channel. The value of $r_{H}/R$ at which the instability occurs grows linearly with $\ell$ as $r_H/R\approx0.1686 + 0.2985\ell$, which means that all GB black holes are unstable at $\alpha = 0.4R^2$.}\label{Ldiverge}
\end{figure}

In the frequency domain we see the instability in the following way. Usually, for the lowest multipole numbers $\ell$ the effective potential is either positive definite or has a tiny negative gap. In this regime, for fixed small values of $\ell$, the eigenfrequencies are damped, though, at some higher $\ell$ there is a sufficiently big negative gap, providing growing modes at some values of black-hole radius $r_{H}$ and $\alpha$. When increasing $\ell$, the parametric region of instability increases, reaching its true limits at asymptotically high $\ell$ (see Fig.~\ref{Ldiverge}). Strictly speaking all frequencies found in the unstable sector are not quasinormal modes, as the whole perturbation approach becomes invalid there, leading to absence of a well-posed initial value problem. However, in \cite{Konoplya:2008yy} it was shown that for arbitrary spherically symmetric black holes quasinormal modes become nonoscillatory, purely damping, regime at the threshold of instability.
In addition, when approaching the threshold of instability, the imaginary part of the fundamental quasinormal frequency approaches zero. This is what we observe in the frequency domain (Fig.~\ref{Ldiverge}).

\begin{figure}
\resizebox{\linewidth}{!}{\includegraphics*{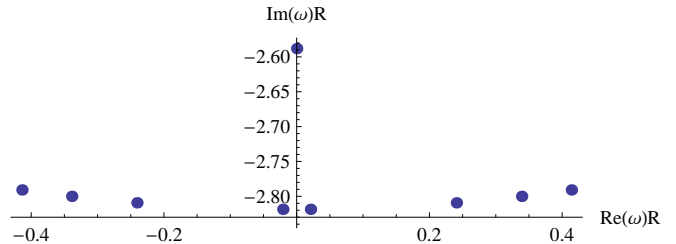}}
\caption{The vector (shear) ``mode'' goes off the imaginary axis for very large negative $\alpha$, i.e. in the regime of instability: $\alpha/R^2 = -2.56, -2.57, -2.58, -2.59, -2.60$ (the real part increases with $\alpha$); $r_H = 10 R$, $\ell =2$, $D=5$.}\label{fig:duplication}
\end{figure}

In \cite{Grozdanov:2016vgg} an interesting phenomenon was found: the vector (shear) modes go off the imaginary axis for higher curvature corrected black holes with planar horizon: initially, purely imaginary modes acquired $\pm$ nonzero real part and, thereby, ``duplicate'', what could indicate the breakdown of the hydrodynamic regime. Here we have observed that this phenomenon occurs only for sufficiently large negative $\alpha$ (Fig.~\ref{fig:duplication}), corresponding to the unstable sector.
%Thus, the would-be breakdown of hydrodynamic description is overshadowed for spherical horizons by the black-hole instability.
In the stable sector, i.e. for small $\alpha$, the dominant vector modes are roughly linear with respect to $\alpha$ and, for example, for $\ell=2, D=5, r_H = 10 R$ obey the formula
$$
\omega \approx -\imo\left(\frac{0.126}{R} - \frac{0.376\alpha}{R^3}\right).
$$

It is known from \cite{Konoplya:2002zu} that in the regime of small black holes, when $r_{H}/R \to 0$ the quasinormal normal modes approach normal modes of empty AdS space-time. When $\alpha$ is nonzero this is evidently not valid anymore owing to the eikonal instability which does not allow the limit $r_{H}/R \to 0$ without requiring simultaneously $\alpha\to 0$.

Another important finding of \cite{Grozdanov:2016vgg} is the presence of a qualitatively new branch of purely imaginary (nonoscillatory) modes
in the scalar channel for GB and $R^4$ theories. Without the GB term, purely imaginary modes for AdS black holes were known only in the vector channel \cite{Konoplya:2003dd}. However, as one can see from Fig. 4 in \cite{Konoplya:2008ix}, similar, scalar, purely imaginary modes appear in the asymptotically flat GB case. Thus, appearance of such modes not only for the planar, but also for the spherical horizon in asymptotically AdS and flat space-times, leads us to the supposition that they might be common for various higher curvature corrected theories. The latter could have observational implications in the asymptotically flat case, when analyzing gravitational waves from black holes in, possibly, alternative, higher curvature corrected theories or from primordial black holes. Indeed, large uncertainty in determination of angular momentum and mass of the resultant black hole leaves open a window for alternative theories \cite{Konoplya:2016pmh}. In $D=5$ it is exactly these purely imaginary modes of the scalar channel that ``drive'' the system to the threshold of instability (Fig.~\ref{Ldiverge}).

Here, for the first time, the eikonal gravitational instability of spherically symmetric Gauss-Bonnet-AdS black holes was shown and exact parametric regions of this instability (given by analytical inequalities) were found. It is shown that even large (relatively the AdS radius $R$) GB-AdS black holes are unstable unless the GB coupling constant $\alpha$ is very small. The instability is accompanied by the breakdown of hyperbolicity. It occurs at much smaller $\alpha$ than those at which the above-mentioned ``duplication'' \cite{Grozdanov:2016vgg} of vector (shear) modes  takes place. Therefore, it is tempting to know what happens in the limit of plane horizon. Taking the limit of the large black hole ($r_H\to\infty$) in (\ref{lowlimit}) and (\ref{uplimit}), we find that planar configurations ($n=3$) should be unstable at
%, i.e. when $r_H\to\infty$. It is easy to see from the previous calculations, that in this regime the eikonal instability for the planar configurations ($n=3$) occurs for
\begin{equation}
\a<-\frac{\sqrt{3}-\sqrt{2}}{2\sqrt{2}}R^2\approx-0.112372R^2,
\end{equation}
and
\begin{equation}
\a>\frac{\sqrt{2}-1}{2\sqrt{2}} R^2\approx0.146447R^2.
\end{equation}

Using the relation between the GB couplings $\lambda_{GB}$ of \cite{Grozdanov:2016vgg} and $\alpha$, used here ($\lambda_{GB}=(\a/R^2)(1-\a/R^2)$), one can find the critical values of  $\lambda_{GB}$ at the onset of instability for the planar geometry of \cite{Grozdanov:2016vgg}. Even if neglecting the nonhyperbolicity of the system, such an instability should develop for the planar GB-AdS black holes at high values of the momentum, which plays the role of the multipole number $\ell$ in the planar case: $k\sim\ell/r_H$ as $r_H\to\infty$. For example, from the fit for the threshold of instability for $D=5$, $\a=0.4R^2$ (Fig.~\ref{Ldiverge}) $r_H/R\approx0.1686 + 0.2985\ell$, we find the critical momentum $k\approx1/0.2985R\approx3.350/R.$ at which the instability should develop. Thus, the instability found here must indicate some limitations of applicability of the GB-AdS backgrounds for holography \cite{Takahashi:2011du}, when $\alpha$ is larger than a critical value, so that the black-hole background does not approach a state of equilibrium.

\section{Eikonal instability and the existence of mass gaps}\label{sec:massgap}

The other aspect of our consideration is related to the claimed stability gap for empty Gauss-Bonnet-AdS space-times \cite{Deppe:2014oua}. In \cite{Deppe:2014oua}, analysis of the critical phenomena for scalar field collapsing in the AdS space-time in the Einstein-Gauss-Bonnet theory was given. It was shown that, if the energy content of the AdS space-time is not sufficiently large, a (small) black hole cannot be formed at a nonzero GB coupling. This fact could be very well explained by the found here linear \emph{instability of any small asymptotically AdS black holes in Gauss-Bonnet theory}: the perturbation cannot ``condense'' into a small black hole, simply because the latter is unstable. Thus, no assumption of stabilization of the pure AdS space by the GB term becomes necessary in such a picture. A further, more accurate study of the critical phenomena in \cite{Deppe:2016dcr} showed that in fact instead of the black-hole formation, a strong and rather chaotic growth of the Ricci-scalar is observed (see p.~15-16 in~\cite{Deppe:2016dcr}). The authors suggested that this means formation of naked singularity, while from our point of view, this also may be manifestation of the instability and nonhyperbolicity of the system.

A similar situation occurs when one considers collapsing massive shells in theories with higher derivatives and implies \emph{asymptotical flatness} of space-time \cite{Frolov:2015bta}. The light black holes do not form in this case either. Thus, it seems that the mass gap in the black hole formation in higher curvature corrected theories does not depend on asymptotic conditions and is intrinsically related to the eikonal instability of small black holes in such theories. It is not excluded that both these phenomena, the mass gap and instability, are consequences of the existence of the fundamental length (or mass) scale in the theory $l_{s}$. If so, then some more fundamental approaches to the problem must also explain the great difference in the critical value of $\alpha$ at thresholds of instability for various numbers of space-time dimensions. The first step in this direction would be a comparison of the critical values of the black-hole mass (at a fixed $\alpha$), which appears in the Choptuik scaling \cite{Deppe:2014oua} and in the thin-shell collapse \cite{Frolov:2015bta}, with the thresholds of instability shown in Figs.~\ref{fig:n3eikonal}~and~\ref{fig:n4eikonal}.

\section{Conclusions}

Here, for the first time, we have shown that Einstein-Gauss-Bonnet-anti-de Sitter black holes are gravitationally unstable at some values of the Gauss-Bonnet coupling constant $\alpha$ and black-hole radius $r_{H}$. The instability develops at high multipole numbers $\ell$ (thus, called \emph{eikonal}) and is accompanied by the breakdown of the well-posedness of the initial value problem. Based on the analysis of effective potentials of the master wave equations we have found exact analytical expressions for the parametric regions of eikonal instability for $D=5$ and $D=6$. In higher space-time dimensions $D$, higher than second curvature corrections must be taken into consideration. We have shown that the found eikonal instability is ``driven'' by the new branch of the purely imaginary modes, which are nonperturbative in $\alpha$ and were recently discussed in the context of perturbations of higher curvature corrected black branes \cite{Grozdanov:2016vgg}. It is shown that as large spherical black holes at values of $\alpha$  higher than the critical are unstable, the same instability must occur for AdS-black holes with planar horizon in the Gauss-Bonnet theory. Finally, we proposed the relation between the existence of the mass gap in the formation of black holes \cite{Frolov:2015bta,Deppe:2016dcr} in higher curvature corrected theories and the observed here instability of small black holes. In order to understand how general the phenomena found here are, it would be interesting to extend out the analysis to the Lovelock and other higher curvature theories.

\acknowledgments{We thank A. Starinets and V. P. Frolov for useful discussions. R.~K. would like to thank the Rudolf Peierls Centre for Theoretical Physics of University of Oxford for hospitality and partial support and the Bridging Grant of the University of T\"ubingen. A.~Z. thanks Conselho Nacional de Desenvolvimento Cient\'ifico e Tecnol\'ogico (CNPq) for support and Theoretical Astrophysics of Eberhard Karls University of T\"ubingen for hospitality.}

\end{document}